\documentclass[aps,twocolumn,showpacs,pra,a4paper,floatfix,nofootinbib]{revtex4-1}%
\usepackage{amsmath}
\usepackage{amssymb}
\usepackage{amsfonts}
\usepackage{fancybox}
\usepackage{eepic}
\usepackage{times}
\usepackage{latexsym}
\usepackage{pifont}
\usepackage{graphicx}
\usepackage{amstext}

\newcommand{\bv}[1]{\mbox{\boldmath $#1$}}
\newcommand{\intensity}[2]{$#1\times10^{#2}\mbox{ W/cm}^{2}$}

\def\jpbo{J. Phys. B: At. Mol. Phys.}	   

\def\pra{Phys. Rev. A}

\def\prl{Phys. Rev. Lett.}

\def\jcp{J. Chem. Phys.}        
\def\md{\mathrm{d}}

\begin{document}

\title{High-order-harmonic generation in benzene with linearly and circularly polarised
laser pulses} 

\author{Abigail Wardlow and Daniel Dundas}

\affiliation{Atomistic Simulation Centre, School of Mathematics and Physics, \\
             Queen's University Belfast, \\
             Belfast BT7 1NN, UK} 
\date{\today}

\begin{abstract}
High-order-harmonic generation in benzene is studied using a mixed quantum-classical
approach in which the electrons are described using time-dependent density
functional theory while the ions move classically. The interaction with both 
linearly and circularly polarised infra-red ($\lambda = 800$ nm) 
laser pulses of duration 10 cycles (26.7 fs) is considered. The effect of
allowing the ions to move is investigated as is the effect of including
self-interaction corrections to the exchange-correlation functional. 
Our results for circularly polarised pulses are compared with previous 
calculations in which the ions were kept fixed and self-interaction corrections
were not included while our results for linearly polarised pulses are compared
with both previous calculations and experiment. We find that even for the short 
duration pulses considered here, the ionic motion greatly influences the harmonic 
spectra. While ionization and ionic displacements are greatest 
when linearly polarised pulses are used, the response to circularly polarised pulses 
is almost comparable, in agreement with previous experimental results.
\end{abstract}

\maketitle

\section{Introduction}
High-order-harmonic generation (HHG) is a highly non-linear process which occurs when
intense, short-duration laser pulses interact with targets such as 
atomic or molecular gases~\cite{bandrauk:2007}. The study of HHG in these systems has attracted much
experimental and theoretical attention in recent years since it represents a
mechanism for the production of attosecond laser pulses and can also be used as
a probe to study how the target atoms or molecules respond during the interaction~\cite{ferre:2015,wong:2011,cireasa:2015}.

Of particular interest is understanding the response of molecules to intense 
laser pulses. This interest arises from the many degrees of freedom that molecules possess which leads to a 
richer set of features and processes when compared to the response of atomic gases. Understanding these
molecular processes is important since they are crucial in many areas of chemistry and biology. 
For example, early simulations showed that charge transfer across the molecule can occur on 
a femtosecond timescale~\cite{remacle:2006}. This opens up the possibility of 
steering electrons in the molecule using attosecond pulses to control the
response and possible fragmentation of the molecule~\cite{krausz:2009}. Due to the extra degrees of 
freedom we expect that molecular HHG will be highly sensitive to changes in the 
molecular dynamics.

Most studies of HHG have focussed on the interaction with linearly polarised pulses.
In the three-step model an electron is liberated by tunnel ionization and propagates as a free
particle in the field before recolliding with the ionic core~\cite{corkum:1993,kulander:1993}. 
For elliptically and circularly polarised light the probability for a direct recollision is greatly reduced -- 
although still present~\cite{rajara:2003,mauger:2010}, and so harmonic generation processes in
these fields are distinctly different. Understanding how elliptically and circularly polarised light
interacts with  molecules is highly important due to the recent experimental and theoretical 
interest in generating circularly polarised attosecond pulses~\cite{ferre:2015} and in studying the
interaction of linearly  and elliptically polarised pulses with chiral 
molecules~\cite{wong:2011,cireasa:2015}. 

A number of theoretical approaches considered molecular HHG in circularly-polarised fields. 
Alon et al~\cite{alon:1998} studied the HHG selection rules that hold when 
symmetric molecules such as benzene interact with circularly polarised laser fields. 
For continuous wave circularly polarised pulses that were polarised parallel to the
molecular plane, they showed that a $6k\pm 1\,\, (k = 0, 1,2, \dots)$ symmetry rule held. 
Averbuch et al~\cite{averbuch:2001} then showed that HHG in this case was
predomnantly due to bound-bound 
transitions. This is distinctly different to bound-continuum transitions that drive HHG in 
the three-step model. Later, Baer et al~\cite{baer:2003} considered the interaction of benzene 
with finite duration, circularly polarised pulses and showed that the $6k\pm 1$ 
selection rule was still evident, even for short pulses. However, a number of 
questions remained unanswered in the study of Baer et al. Firstly, their calculations used 
time-dependent density functional theory (TDDFT)~\cite{runge:1984} in which many-body effects 
were treated at the level of the local density approximation. It is well known that such an 
approximation includes self-interaction errors. In particular,  the exchange-correlation 
potential does not have the proper asymptotic form meaning that the ionization potential of 
the molecule is underestimated. Secondly, their calculations considered the ions to be fixed 
in space. As mentioned above, HHG is highly sensitive to changes in the molecular response 
and so the effect of allowing the ions to move, even for such short duration pulses, has not 
been fully explored.

Describing HHG in complex molecules is theoretically and computationally demanding. The number 
of degrees of freedom in the problem together with the duration of the laser pulses and the 
length scales required to describe recolliding electrons means that solution of the 
time-dependent Schr\"{o}dinger equation is impractical for all but the simplest one- and 
two-electron molecules~\cite{becker:2008,kono:1997,lein:2003,dundas:2003,lein:2003,%
bandrauk:2004,saenz:2010,kjeldsen:2006,morales:2014,hou:2012,yuan:2012}. For more complex 
molecules, approximations are required. One successful technique for studying HHG in complex 
molecules is the strong field approximation (SFA) in which either single or multiple ionic
channels can be considered~\cite{spanner:2009,smirnova1:2009,cireasa:2015}. While this method 
has allowed HHG in general elliptically polarised laser pulses to be investigated, it does 
suffer from the drawback that only a small number of channels are generally included in the 
calculation. This can become an issue, depending on the molecule under investigation, if other 
channels become important. Indeed, such a situation has been observed in other ab initio 
calculations in which multiple electrons can contribute to the molecular
response~\cite{chu:2004,dundas:2004b,bandrauk:2011,petretti:2010}. Another successful technique 
for studying HHG is quantitative rescattering (QRS) theory~\cite{le:2009,lin:2010}. In this approach,
the induced dipoles can be expressed as the product of a recolliding electronic wavepacket with a
photorecombination cross section. Multiple orbitals can be incorporated within QRS and the method
has successfully applied to HHG in complex molecules~\cite{jin:2011a,wong:2013,le:2013a,le:2013b}. Like
the strong-field approximation, the influence of multiple orbitals in QRS can only be studied if they 
have been included from the outset.
One ab initio approach that is widely used for studying laser molecule interactions is 
TDDFT. While it has mainly been used to study the electronic 
response~\cite{telnov:2009,heslar:2011,bandrauk:2011,petretti:2010,baer:2003}
some implementations combine TDDFT with a classical description of the ionic 
motion~\cite{kunert:2003,uhlmann:2003,dundas:2004,calvayrac:2000,
castro:2004,dundas:Jchemphys:2012}. To date, very few implementations combine the
treatment of ionic motion with approximations to the exchange-correlation 
functional that include self-interaction corrections~\cite{wopperer:2012,crawford:2014}. 
However, these calculations have generally considered the interaction with high-frequency 
laser pulses. As far as we are aware, no calculations of HHG in infra-red (IR) pulses have 
been carried out before that combine moving ions with self-interaction corrections.

In this paper we study HHG in benzene using both circularly and linearly polarised IR laser 
pulses. We use a mixed quantum-classical approach in which electrons are treated using TDDFT
while the ions are described classically. In our treatment, self-interaction corrections to
the exchange-correlation functional are included. The paper is arranged as follows. 
In Sec.~\ref{sec:theory} we describe our approach and give details of the numerical methods 
and parameters used in our simulations. In Sec.~\ref{sec:results} our results are presented. 
Firstly, we consider benzene interacting with circularly-polarised pulses. The effect of 
including self-interaction corrections and allowing the ions to move is investigated. 
Secondly, we show how the response of the molecule changes when linearly polarised pulses are used. 
In particular, we show that the response of the ions is slightly greater for 
linearly polarised pulses. Finally, in Sec.~\ref{sec:conclusions} we present
our conclusions and possible directions for future work are highlighted.

Unless otherwise stated, atomic units are used throughout.

\section{Theoretical Approach}
\label{sec:theory}

In our calculations we consider quantum mechanical electrons and classical ions 
interacting with intense, ultra-short laser pulses in a method known as non-adiabatic 
quantum molecular dynamics (NAQMD). The implementation of this method, in a code called 
EDAMAME (Ehrenfest DynAMics on Adaptive MEshes), is described in more detail in 
Ref.~\cite{dundas:Jchemphys:2012}. We now briefly 
describe how it is applied to the calculations considered here.

We consider $N_n$ classical ions where $M_k$, $Z_k$ and $\bv{R}_k$ denote respectively 
the mass, charge and position of ion $k$. Additionally $\bv{R} = \{\bv{R}_1, \dots, \bv{R}_{N_n}\}$. 
Time dependent density functional theory (TDDFT)~\cite{runge:1984} is used to model the electronic 
dynamics. Neglecting electron spin effects, the time-dependent electron density can be written in 
terms of $N$ time-dependent Kohn-Sham orbitals, $\psi_{i} (\bv{r},t)$, as
\begin{equation}
n(\bv{r},t) =  2\sum_{j=1}^{N} \left | \psi_{j } (\bv{r},t)  \right |^2.
\end{equation}
These Kohn-Sham orbitals satisfy the time dependent Kohn-Sham equations (TDKS)
\begin{align}
i\frac{\partial }{\partial t}\psi_{j} (\bv{r},t) = 
\Biggr [ & 
-\frac{1}{2} \nabla^2 + V_{\mbox{\scriptsize H}}(\bv{r}, t)+
V_{\mbox{\scriptsize ext}}(\bv{r}, \bv{R}, t)
\nonumber\\
& 
+ V_{\mbox{\scriptsize xc}}(\bv{r},  t) \Biggr] \psi_{j} (\bv{r},t) = 
H_{\mbox{\scriptsize ks}}\psi_{j} (\bv{r},t)
.
\label{eq:tdks}
\end{align}
In Eq.~(\ref{eq:tdks}), $V_{\mbox{\scriptsize H}}(\bv{r}, t)$ is the Hartree potential, 
$V_{\mbox{\scriptsize ext}}(\bv{r}, \bv{R}, t)$ is the external potential, and 
$V_{\mbox{\scriptsize xc}}(\bv{r},  t)$ is the exchange-correlation  potential. Both the Hartree and
exchange-correlation potentials are time-dependent due to their functional dependence on the
time-dependent density.
The external potential accounts for both electron-ion interactions and the interaction of the 
laser field with the electrons and can be written as
\begin{equation}
V_{\mbox{\scriptsize ext}}(\bv{r}, \bv{R}, t) =
V_{\mbox{\scriptsize ions}}(\bv{r}, \bv{R}, t) +
U_{\mbox{\scriptsize elec}}(\bv{r}, t).
\end{equation}
The calculations presented in this paper will consider the response of benzene to IR laser 
pulses. In that case the laser will predominately couple to the valence electrons and so we
will only consider their response. The electron-ion interactions, 
$V_{\mbox{\scriptsize ions}}(\bv{r}, \bv{R}, t)$, are described with Troullier-Martins 
pseudopotentials~\cite{Troullier:1991} in the Kleinman-Bylander form~\cite{Kleinman:1982}. 
All pseudopotentials were generated using the APE (Atomic Pseudopotentials 
Engine)~\cite{Oliveira:2008}. 

Both length and velocity gauge descriptions of the electron-laser 
interaction, $U_{\mbox{\scriptsize elec}}(\bv{r}, t)$, can be
used by EDAMAME. We use the length gauge description here as it the simpler to implement
and major differences between the gauges have not previously been observed~\cite{dundas:Jchemphys:2012}. 
Working within the dipole approximation we then have  
\begin{equation}
U_{\mbox{\scriptsize elec}}(\bv{r},t) = U_L (\bv{r},t) = \bv{r} \cdot \bv{E}(t).
\end{equation}

\begin{table*}[t]
\begin{tabular*}{0.7\textwidth}{@{\extracolsep{\fill}} cccc} \hline\hline
\multicolumn{1}{c}{} &
\multicolumn{2}{c}{Equilibrium bond lengths (a.u.)} &
\multicolumn{1}{c}{Ionization Potential (a.u.)}\\
\multicolumn{1}{c}{} &
\multicolumn{1}{c}{C--C bond length} &
\multicolumn{1}{c}{C--H bond length} &
\multicolumn{1}{c}{} \\
\multicolumn{1}{c}{LDA-PW92} & 2.622 & 2.066 & 0.243 \\
\multicolumn{1}{c}{L-ADSIC} & 2.541 & 2.138 & 0.345 \\
\multicolumn{1}{c}{Experimental} & 2.644 & 2.081 & 0.340\\
\hline\hline
\end{tabular*}
\caption{Static properties of benzene. The equilibrium C-C and C-H bond lengths and ionization 
potential calculated using LDA-PW92 and L-ADSIC are compared with experimental values. The ionization 
potential is estimated from the Highest occupied molecular orbital (HOMO) orbital energy. Other calculation 
parameters are detailed in the text. We see that the C-C bond lengths are underestimated with L-ADSIC while 
the C-H bond lengths are overestimated. However, the ionization potential is much more accurate using 
L-ADSIC. Experimental bond lengths are taken from~\protect\cite{meijer:1996} while the ionization potential 
is taken from~\protect\cite{sharifi:2007}}
\label{tab:table1}
\end{table*}

In the calculations presented in the paper we consider benzene interacting with both linearly  and 
circularly polarised pulses. In the case of linearly polarised light, we take the polarisation direction 
to be along the
$x$ axis. In that case, the vector potential has the form 
\begin{equation}
\bv{A}(t)= A_0f(t)\cos(\omega_L t + \phi )\bv{\hat{e}}_x.
\end{equation}
Here $A_0$ is the peak value of the vector potential, $\omega_L$ is the laser frequency, $\phi$ is the 
carrier-envelope phase and $f(t)$ is the pulse envelope which we model as
\begin{equation}\label{eq:pulseenvlope}
f(t) = \left\{\begin{matrix}\displaystyle \sin^2 \left( \frac{\pi t}{T}\right ) & 0 \leq  t \leq  T\\[0.4cm] 
0 & \textup{otherwise} \end{matrix}\right.,
\end{equation}
where $T$ is the duration of a pulse. From this, the electric vector is
\begin{equation}
\bv{E}(t)= E(t)\bv{\hat{e}}_x,
\end{equation}
where 
\begin{equation}
E(t) = E_0 f(t) \sin(\omega_L t + \phi) - 
\frac{E_0}{\omega_L}\frac{\partial f}{\partial t} \cos (\omega_L t + \phi),
\end{equation}
and where $E_0$ is the peak electric field strength.

For circularly polarised light, we consider the pulse to be polarised in the $x-y$ plane so that the 
vector potential is given by 
\begin{align}
\bv{A}(t) & = A_0 f(t) \Bigr(\cos (\omega_L t +\phi) \bv{\hat{e}}_x +
                             \sin (\omega_L t +\phi)\bv{\hat{e}}_y\Bigr)\nonumber\\
          & = A_x(t)\bv{\hat{e}}_x + A_y(t)\bv{\hat{e}}_y.
\end{align}
The pulse envelope takes the form given in Eq.~(\ref{eq:pulseenvlope}). In that case
the electric field is given by
\begin{equation}
\bv{E}(t)= E_x(t)\bv{\hat{e}}_x + E_y(t)\bv{\hat{e}}_y,
\end{equation} 
where
\begin{equation}
E_x(t) = \frac{E_0}{\sqrt{2}} f(t) \sin(\omega_L t + \phi) - 
\frac{E_0}{\sqrt{2}\omega_L}\frac{\partial f}{\partial t} \cos (\omega_L t + \phi),
\end{equation} 
and
\begin{equation}
E_y(t) = -\frac{E_0}{\sqrt{2}} f(t) \cos(\omega_L t + \phi) - 
\frac{E_0}{\sqrt{2}\omega_L}\frac{\partial f}{\partial t} \sin (\omega_L t + \phi).
\end{equation} 

The exchange-correlation potential, $V_{\mbox{\scriptsize xc}}(\bv{r},  t)$,
accounts for all electron-electron interactions. The exact form of the exchange-correlation 
action functional is unknown and so must be approximated. In this paper, we consider two adiabatic 
approximations to the exchange-correlation potential. 
In the first approximation, the local density approximation (LDA) incorporating the Perdew-Wang 
parameterization of the correlation functional is used~\cite{Perdew:1992}; we will refer to this method 
as LDA-PW92. The LDA derives from describing the electrons as a homogeneous electron gas. Although 
LDA is widely used, it has self-interaction errors which means that it does not have the 
correct long range behaviour. In particular, the exchange-correlation potential decays exponentially 
instead of Coulombically, meaning that ionization potentials are underestimated. In the second 
approximation, the LDA-PW92 functional is supplemented by the average density 
self-interaction correction (ADSIC)~\cite{Legrand:2002}; we will refer to this method as L-ADSIC. 
L-ADSIC aims to correct the self-interaction errors through use of an orbital-independent term involving 
the average of the total electron density. A major benefit of L-ADSIC is that the exchange-correlation potential can 
be obtained as the functional derivative of an exchange-correlation functional,
meaning that the forces acting on ions can be derived from the Hellmann-Feynman theorem.

We solve the Kohn-Sham equations numerically using finite difference techniques
on a three dimensional grid. All derivative operators are approximated using 9-point
finite difference rules and the resulting grid is parallelised in 3D. The following mesh parameters were used
for the calculations presented in this paper. In those simulations where circularly-polarised 
pulses were used, the grid extents were $|x| \leq 76.8\,a_0$, $|y| \leq 76.8\, a_0$ and 
$|z| \leq 48.8\, a_0$ while for 
linearly-polarised pulses the extents were $|x| \leq 104.8\, a_0$, $|y| \leq 62.8\, a_0$ and 
$|z| \leq 62.8\, a_0$. A larger grid extent along the laser polarisation direction is used whenever the
response to linearly-polarised pulses is considered. This is because we need to handle recolliding 
wavepackets that travel predominantly in this direction. Grid spacings of $0.4\, a_0$ were used for all coordinates.

The ionic dynamics are treated classically using Newton's equations of motion. For ion $k$ we
have
\begin{align}
M_k \ddot{R}_k = &- 
\int n (\bv{r},t) \frac{\partial H_{\mbox{\scriptsize ks}}}{\partial \bv{R}_k} 
\md\bv{r} \nonumber \\
& -\frac{\partial}{\partial \bv{R}_k} \Bigr(V_{nn} (\bv{R}) + Z_k \bv{R}_k\cdot \bv{E}(t)\Bigr),
\label{eq:neom}
\end{align}
where $V_{nn} (\bv{R})$ is the Coulomb repulsion between the ions and $Z_k \bv{R}_k\cdot \bv{E}(t)$ 
denotes the interaction between ion $k$ and the laser field. 

We need to propagate both the TDKS equations, Eq.~(\ref{eq:tdks}), and ionic equations of motion, Eq.~(\ref{eq:neom}),  
in time. We propagate the TDKS equations using an 18th-order unitary Arnoldi 
propagator~\cite{dundas:2004, Arnoldi:1951, Smyth:1998}. For the ionic equations of motion, we use 
the velocity-Verlet method. Converged results are obtained for a timestep of $0.2$ a.u.

As a simulation progresses, electronic wavepackets can travel to the edge of the spatial grid, reflect  
and travel back towards the molecular centre. These unphysical reflections can be overcome using 
a wavefunction splitting technique that removes wavepackets that reach the edge of the 
grid~\cite{Smyth:1998}. The splitting is accomplished using a mask function, $M(\bv{r})$, 
that splits the Kohn-Sham orbital, $\psi_j(\bv{r},t)$, into two parts. One part, 
$M (\bv{r})\psi_j(\bv{r},t)$, is located near the molecule and is associated with non-ionized wavepackets. The
other part, $\left \{ 1-M(\bv{r}) \right \} \psi_j(\bv{r},t)$, is located far from the molecule and is
associated with ionizing wavepackets: this part is discarded in our calculations. The point at which we apply this splitting must be 
be chosen carefully to ensure only ionizing wavepackets are removed. 
We write the mask function in the form 
\begin{equation}
M(\bv{r}) = M_x(x)M_y(y)M_z(z).
\end{equation}
If we consider the $x$ component we can write
\begin{equation}\label{eq:mx}
M_x(x)=\left\{\begin{matrix}
1  & \left | x \right | \leq  x_m\\[0.4cm] 
1 - \alpha \left ( \left | x \right | -x_m \right )^5 & \left | x \right | > x_m
\end{matrix}\right. ,
\end{equation}
where
\begin{equation}
\alpha = \frac{1- M_f}{(x_f - x_m)^5}.
\end{equation}
Here $x_m$ is the point on the grid where the mask starts, $x_f$ is the maximum extent of the grid in 
$x$ and $M_f$ is the value that we want the mask function to take at the edges of the grid. 
Similar descriptions are used for $M_y(y)$ and $M_z(z)$. 

Within TDDFT ionization is a functional of the 
electronic density. The exact form of this functional is unknown and so most measures of 
ionization are obtained using geometric properties of the time-dependent Kohn-Sham
orbitals~\cite{ullrich:2000}. In this approach, bound- and continuum-states are separated into 
different regions of space through the introduction of an analysing box. In principal the continuum
states occupy the regions of space when the wavefunction splitting is applied and hence the 
reduction in the orbital occupancies provides a measure of ionization. 

\section{Results}
\label{sec:results}

\begin{figure}
\centerline{\includegraphics[width=6cm,viewport=20 51 385 322]{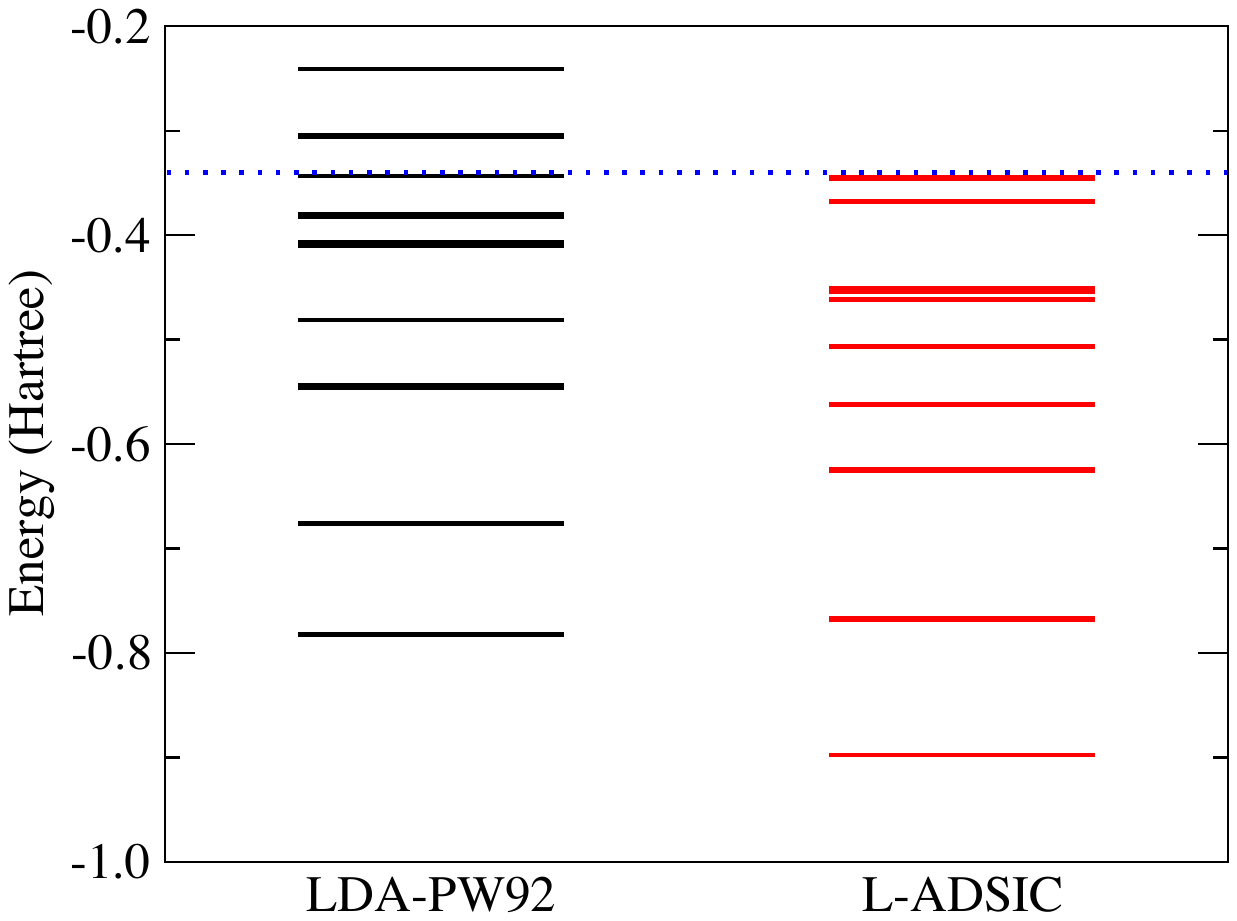}
}
\caption{Kohn-Sham orbital energies of benzene obtained from relaxed geometries using both
LDA-PW92 and L-ADSIC approximations to the exchange-correlation functional. The dotted line is
plotted at the negative of the experimental ionization potential of benzene. We see that 
using L-ADSIC the HOMO orbital energy gives a good approximation to this ionization potential.}
\label{fig:figure1}
\end{figure}

The starting point for our calculations is to obtain the ground state of benzene. We do this by taking a trial guess for 
the geometry from the NIST Chemistry WebBook~\cite{nist}. Relaxing this initial structure using the different 
exchange-correlation potentials gives the equilibrium properties presented in Table~\ref{tab:table1}. We see from these 
structural properties that the equilibrium geometry is well reproduced using LDA-PW92. Using L-ADSIC we see that the C--C 
bond lengths are underestimated while the C--H bond lengths are overestimated. This is similar to what is 
found using other ADSIC calculations~\cite{Legrand:2002}. Additionally we see that the ionization potential is better 
represented using L-ADSIC than that obtained using LDA-PW92. Indeed, if we plot the Kohn-Sham orbital energies, 
as shown in Fig.~\ref{fig:figure1}, we see that L-ADSIC gives a better representation of the electronic
structure than LDA-PW92. In particular, we see that several Kohn-Sham orbitals are much more loosely bound
when using LDA-PW92.

\begin{figure*}
\hfill
\begin{center}
\begin{minipage}{9.5cm}
\includegraphics[width=9.5cm,viewport=10 180 545 655]{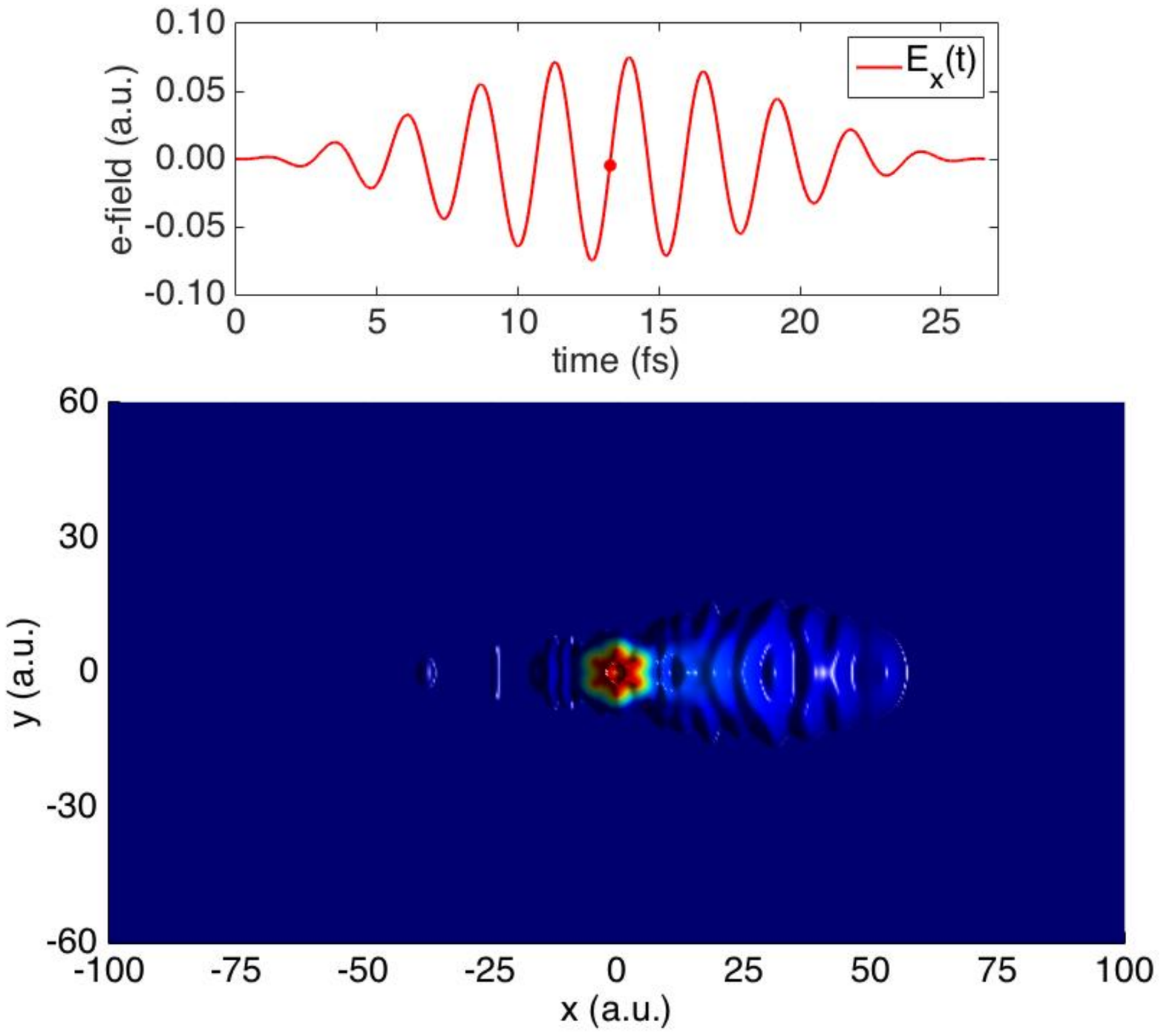}
\centerline{(a)}
\end{minipage}
\hfill
\begin{minipage}{7.0cm}
\includegraphics[width=7.0cm,viewport=45 135 480 700]{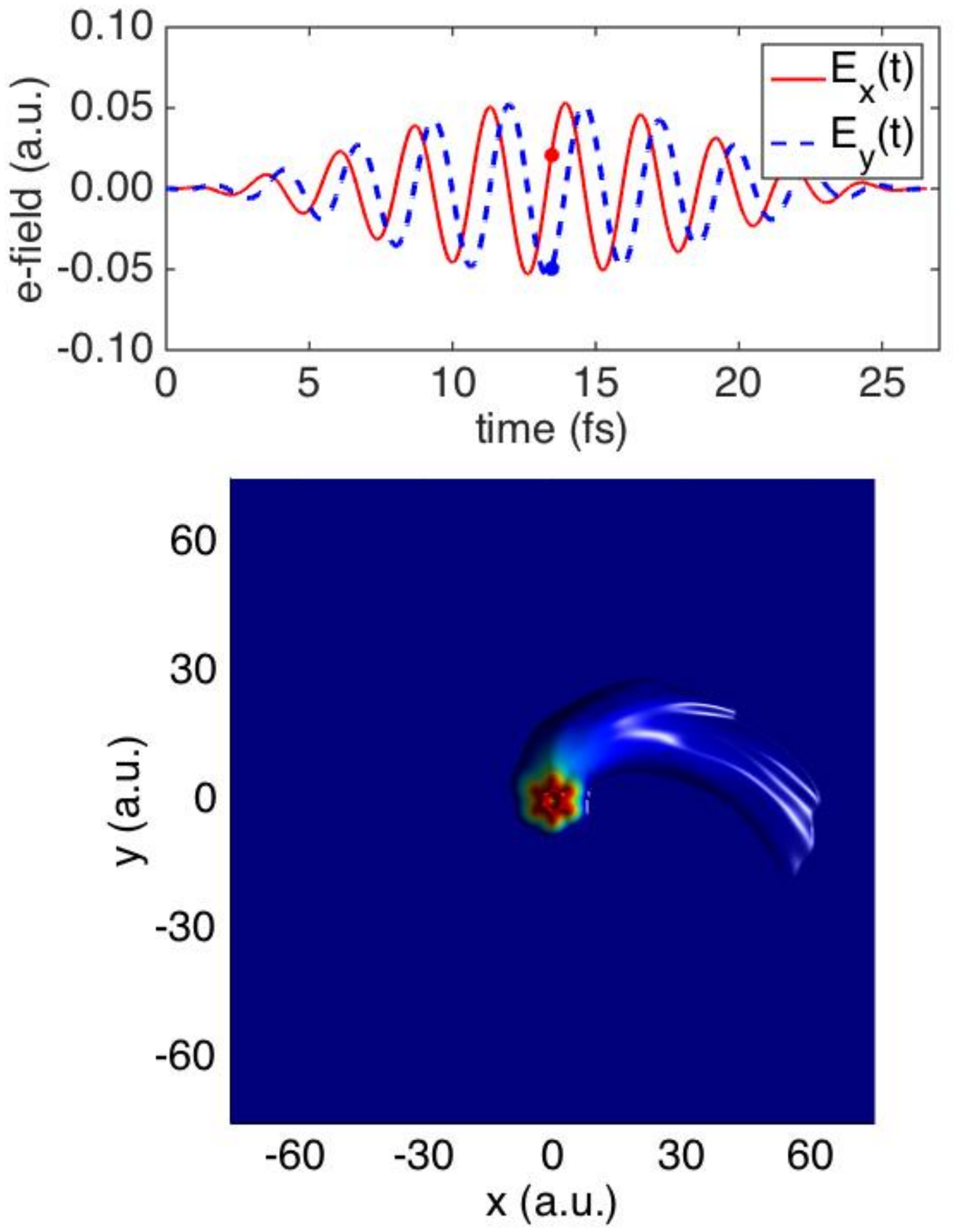}
\centerline{(b)}
\end{minipage}
\hfill
\end{center}
\caption{Snapshots of the electronic density of benzene during interaction with (a) a 10-cycle linearly-polarised
pulse having a peak intensity of \intensity{2.0}{14} and wavelength of 800 nm and (b) a 10-cycle circularly-polarised
pulse having a peak intensity of \intensity{2.0}{14} and wavelength of 800 nm. In both cases the L-ADSIC
approximation has been used and the ions are allowed to move. The densities presented are obtained by integrating the 
full 3D density over the $z$ coordinate. The densities are plotted on a logarithmic scale and the 
lowest density has been clamped to $10^{-4}$.}
\label{fig:figure2}
\end{figure*}

We are now in a position to consider benzene interacting with intense laser pulses. The first question we can
ask is how does the molecule respond to circularly- and linearly-polarised light? We visualise both these situations
in Fig.~\ref{fig:figure2}. In both simulations the L-ADSIC exchange-correlation potential is used and the ions are allowed 
to move. The densities presented are obtained by integrating the full 3D density over the $z$ coordinate. 
Fig.~\ref{fig:figure2}(a) presents a snapshot of the electronic density of benzene during its interaction with a 10-cycle 
linearly-polarised pulse having a peak intensity of $I = $ \intensity{2.0}{14} and wavelength of $\lambda =$ 800 nm. The laser 
is polarised along the $x$-axis. We clearly see that the electronic response is
predominantly along this  direction. In the frame shown the electron is being forced along the positive $x$-direction. 
We see electrons streaming outwards in this direction. At small negative $x$ values we can also clearly see ionized 
electrons recolliding with the parent molecule. For circular polarisation, the situation is markedly different. 
Here we see electrons spiralling outwards as the molecule responds to the field. While most electrons never return 
to the parent core, we see some evidence of electron wavepackets ionizing from one atomic centre and subsequently recombining 
with at an adjacent atomic centre: this is more evident when we consider a movie of the evolving density.
Animations of both simulations are presented in the supplementary material.

Now we know that the electronic response is markedly different we can consider ionization, harmonic generation and the 
response of the ions to the different laser pulses.
For HHG, we calculate the spectral density, $S_k(\omega)$, along the direction 
$\hat{\bv{e}}_k$ from the Fourier transform 
of the dipole acceleration~\cite{burnett:1992} 
\begin{equation}
S_k(\omega) = \left | \int e^{ i \omega t } \,\hat{\bv{e}}_k \cdot 
\ddot{\bv{d}}(t)\,\md t \right |^2,
\end{equation}
where $\ddot{\bv{d}}(t)$ is the dipole acceleration given by 
\begin{equation}
\ddot{\bv{d}}(t) = - \int n(\bv{r},t) \bv{\nabla} H_{\mbox{\scriptsize ks}} \md\bv{r}.
\end{equation}

\subsection{Molecular response to circularly-polarised light}
\begin{figure}[t]
\centerline{\includegraphics[width=8cm,viewport=18 42 594 529]{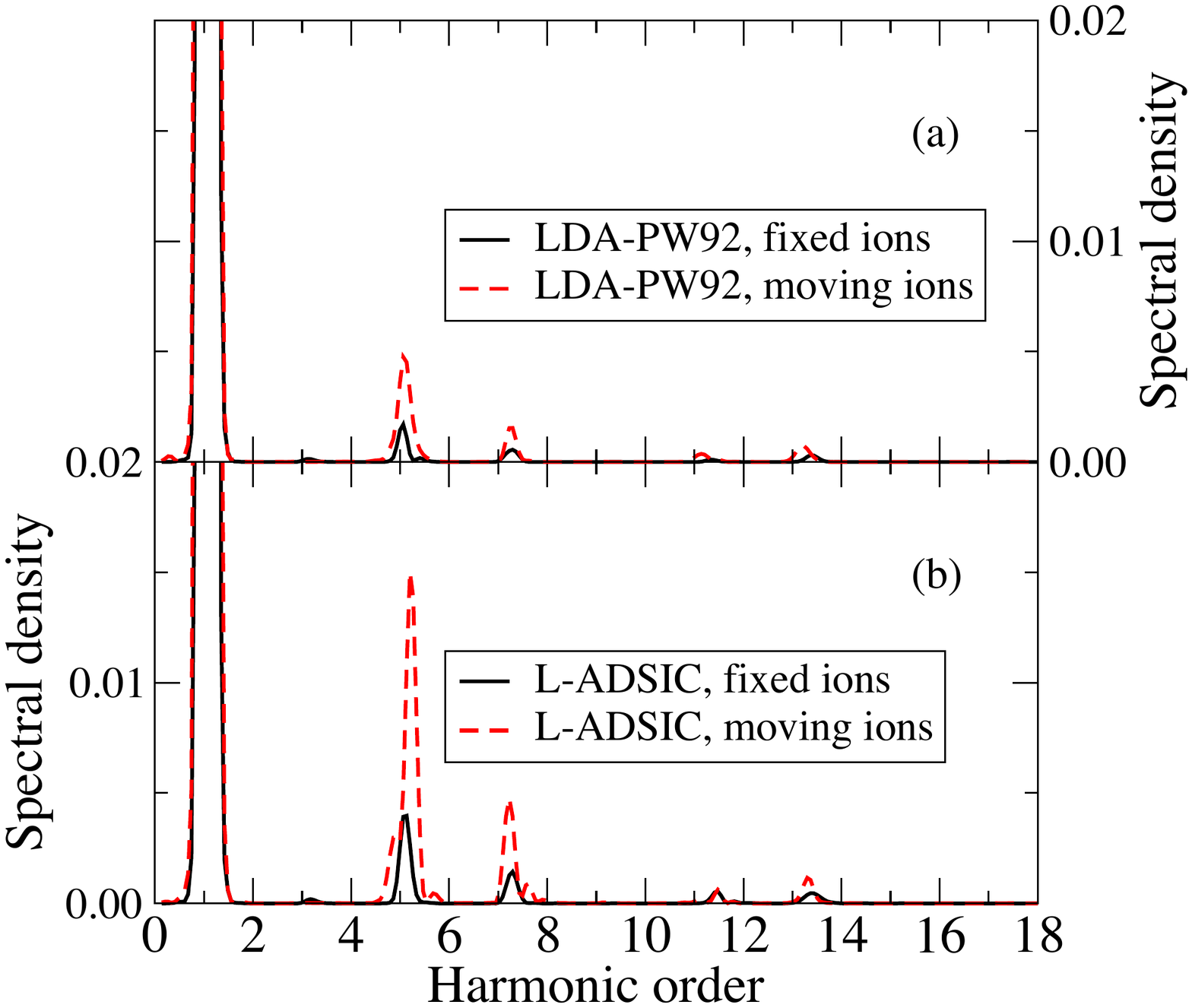}}
\caption{Harmonic generation in benzene during interaction with a 10-cycle circularly-polarised 
laser pulse having a peak intensity $I =$ \intensity{3.5}{14} and wavelength $\lambda =$ 800 nm. 
We compare how the response changes when the ions are kept fixed and allowed to move. Plot (a) shows 
the calculation made using the LDA-PW92 exchange-correlation functional and plot (b) shows the calculation made using
the L-ADSIC exchange-correlation functional.}
\label{fig:figure3}
\end{figure}

Baer et al~\cite{baer:2003} studied HHG in benzene with TDDFT using the LDA-PW92
approximation to the exchange-correlation functional. In their work the ions were kept fixed
in space and the harmonic response was studied for a range of laser intensities and pulse lengths.
Here we are interested in how our results are altered by using a more accurate exchange-correlation
functional and allowing the ions to move. Thus we initially consider the interaction with a
10-cycle laser pulse of wavelength $\lambda=$ 800 nm and peak intensity $I =$ \intensity{3.5}{14}. To compare 
with Baer et al we will present harmonic spectra for the response in the $x$ direction, i.e. $S_x(\omega)$.
Figure~\ref{fig:figure3} compares the harmonic spectra when the ions are kept fixed and when they are allowed 
to move. Both exchange-correlation functionals have been used. 
Even for these short-duration pulses, we see that the $6k\pm 1, (k = 0, 1,\dots)$ selection rule 
holds in all cases, i.e. the 3rd, 9th and 15th harmonics are suppressed. In all cases we can compare the relative
strengths of the harmonics obtained. These are presented in the first part of Table~(\ref{tab:table2}) where 
our results are compared with those of Baer et al~\cite{baer:2003}. Our results show that the relative strengths 
change greatly depending on the exchange-correlation functional used and whether or not the ions are allowed to 
move. We note two points of interest from Fig.~\ref{fig:figure3}. Firstly, for both 
exchange-correlation functionals we see that the harmonic intensities increase whenever the ions are allowed to move.
Secondly, the harmonic intensities are larger for the L-ADSIC calculations than for the LDA-PW92 calculations. 

\begin{figure}[t]
\centerline{\includegraphics[width=8cm,viewport=14 43 564 496]{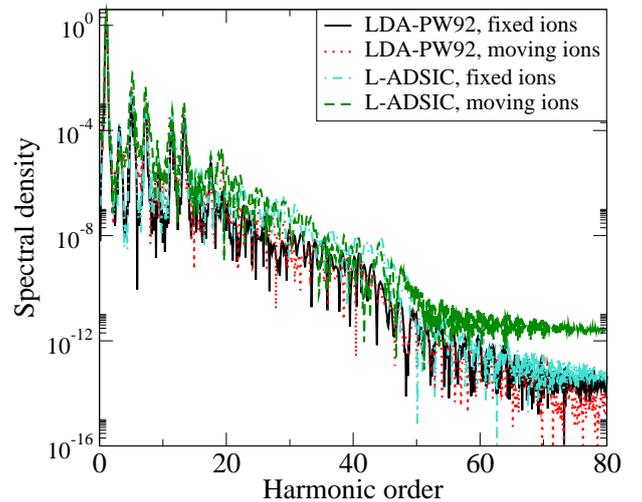}}
\caption{Harmonic generation in benzene  during interaction with a 10-cycle circularly-polarised laser pulse 
having a peak intensity $I =$ \intensity{3.5}{14} and wavelength $\lambda =$ 800 nm. Logarithmic plots of the 
four spectra in Fig.~\protect\ref{fig:figure3} are presented here. We can clearly see the formation of a 
plateau region. Based on the classical cut-off formula, the maximum plateau harmonic for this laser intensity 
and wavelength would be harmonic 49. We can clearly see that the intensities of the plateau harmonics increase 
when the L-ADSIC functional is used. In addition, for each functional, we see that the harmonic intensities are 
greater when the ions are allowed to move.}
\label{fig:figure4}
\end{figure}

In the results of Baer et al~\cite{baer:2003}, it was noted that a secondary plateau was observed in the harmonic
response. It was suggested that this plateau is created due to the finite-width laser frequency. 
In Fig.~\ref{fig:figure4} we present logarithmic plots of the harmonic spectra for the four calculations considered in
Fig.~\ref{fig:figure3}. In all cases we see a secondary plateau. A number of observations can be made
from these results. Firstly, as in Fig.~\ref{fig:figure3}, the intensities of the harmonics in this secondary plateau are 
greater for L-ADSIC calculations than for the LDA-PW92 results. Secondly, the harmonic intensities in the secondary plateau increase 
when the ions are allowed to move. 

Averbukh et al~\cite{averbuch:2001} showed that, for the laser pulses considered here, HHG in benzene is predominantly due to 
bound-bound transitions rather than bound-continuum transitions. However, bound-continuum transitions should still be
present. For example, recollisions during the interaction of circularly polarised 
light with atoms has already been considered by several authors~\cite{rajara:2003,mauger:2010} while in 
Fig.~\ref{fig:figure2}(b) we see direct evidence of electron recollisions. 
We are able to explain the results observed in Figs~\ref{fig:figure3} and~\ref{fig:figure4} in terms of bound-bound and
bound-continuum transitions. Referring to Fig.~\ref{fig:figure1}, we see that, as well as giving a 
better description of the electronic structure of benzene, the L-ADSIC approximation leads to many more resonant 
transitions between Kohn-Sham states for the laser wavelength considered. The increase in 
the primary plateau harmonics using L-ADSIC can therefore be explained in terms of enhanced bound-bound transitions arising from the 
more accurate electronic structure. Furthermore, we believe that increases in the harmonic intensities for moving ions, particularly
in the secondary plateau, is primarily due to bound-continuum transitions. In this case, a potential mechanism is 
electrons that tunnel-ionize from one atomic centre recolliding with an adjacent atomic centre as they spiral in the 
field. If the ions have sufficient displacement from their equilibrium positions, then the probability for such 
recollisions to occur will increase. Based on the classical cut-off formula for HHG, the maximum plateau harmonic for the laser
parameters considered here should be 49. This is in good agreement with the results in Fig.~\ref{fig:figure4}.
Obviously, such recollisions will also enhance the primary plateau harmonics, as we observe.

\begin{figure*}[t]
\centerline{
\begin{minipage}{8.0cm}
\centerline{\includegraphics[width=8cm,viewport=28 37 565 498]{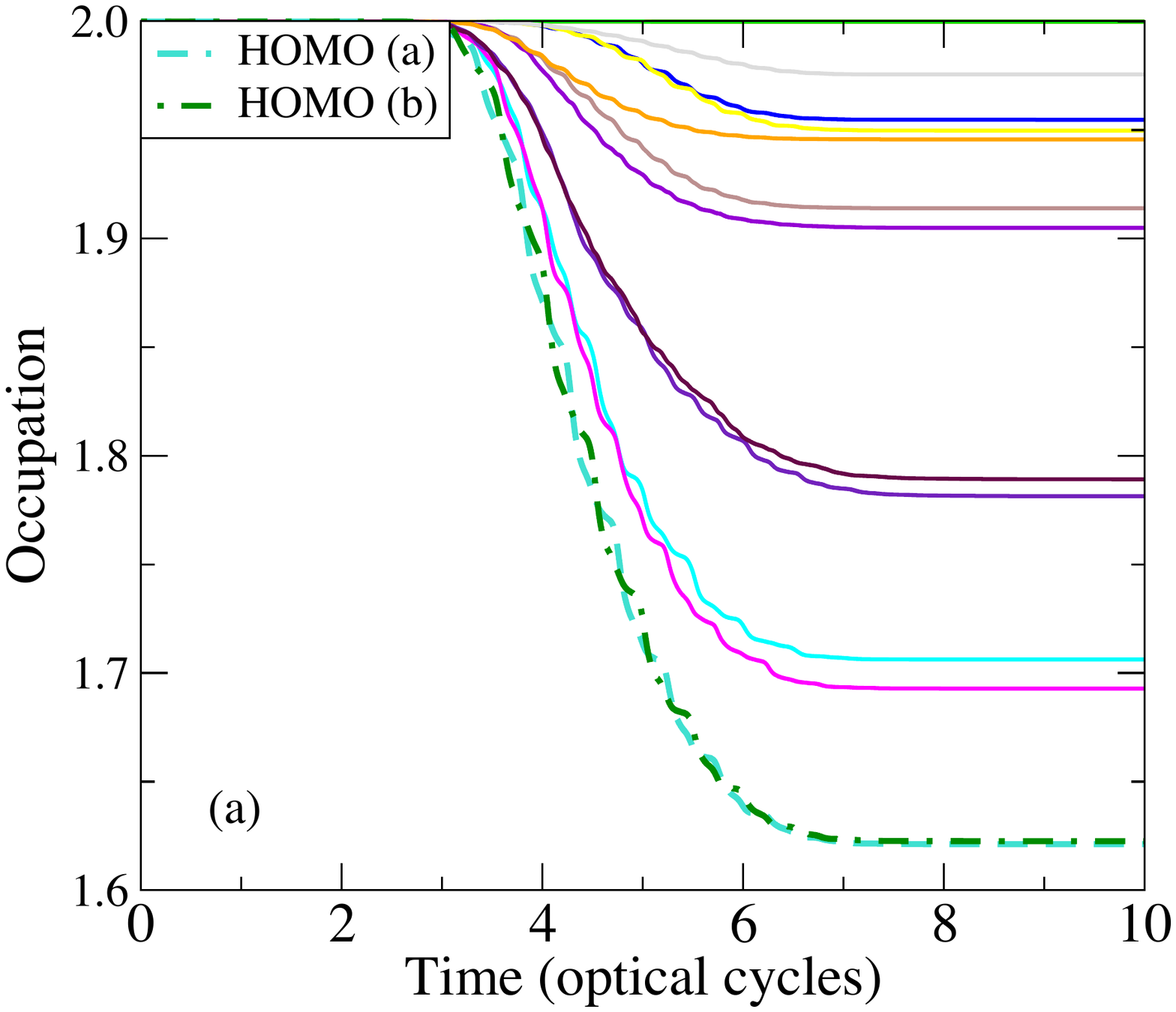}}
\end{minipage}
\hfill
\begin{minipage}{8.0cm}
\centerline{\includegraphics[width=8cm,viewport=28 37 565 499]{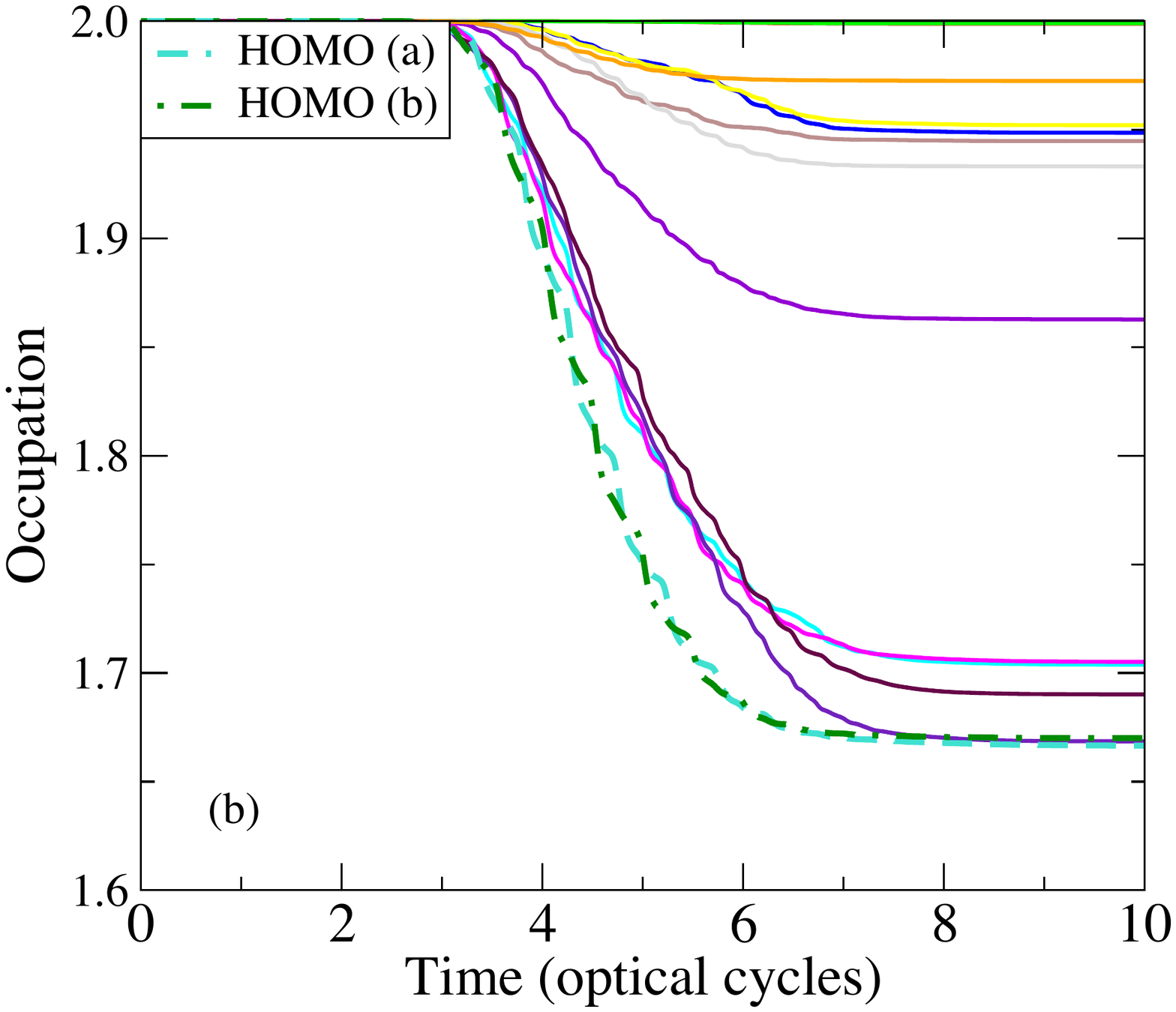}}
\end{minipage}
} 
\caption{Populations of the Kohn-Sham orbitals of benzene during the interaction with a 10-cycle circularly-polarised laser 
pulse of wavelength $\lambda =$ 800 nm and intensity $I =$ \intensity{3.5}{14}. In (a) the ions are kept fixed during the
simulation while in (b) the ions are allowed to move. All calculations were performed using the L-ADSIC exchange-correlation
functional. In the simulations both the molecule and the polarisation of the laser lie in the $x-y$ plane. For clarity we only
label the two forms of the doubly degenerate HOMO orbital: these are referred to as HOMO(a) and HOMO(b). The reduction of the population due to the wavefunction splitting
method provides a measure of the amount of ionization. By the end of the pulse the total electron depletion is 2.06 for the moving ion
calculation, compared to 1.98 for the fixed-ion calculation.}
\label{fig:figure5}
\end{figure*}

In order to study how the electrons and ions respond to the laser pulse, 
we can consider the response of the atom trajectories and the
depletion of the Kohn-Sham orbitals (a measure of the amount of ionization). Fig.~\ref{fig:figure5} presents the orbital
response for the L-ADSIC calculations for benzene interacting with 10-cycle laser pulse of wavelength $\lambda=$ 800 nm 
and peak intensity $I =$ \intensity{3.5}{14}. Fig.~\ref{fig:figure5}(a) presents the results for fixed ions while Fig.~\ref{fig:figure5}(b) presents
the results for moving ions. The reduction in orbital occupation gives a measure of the amount of ionization. For clarity, in these plots
we only label the doubly-degenerate HOMO orbital [HOMO(a) and HOMO(b)]. In the case of fixed ions, we see that the HOMO orbital has the greatest response. 
However, we see that the other more tightly bound orbitals also respond significantly to the field. Hence, multielectron
effects are important in describing the response of the molecule. When the ions are allowed to move the response is 
different. We now see that the total depletion is slightly greater, i.e. more ionization has occurred. Additionally, the
response of the orbitals is also different. In this case some of the more tightly bound orbitals have a response similar to the
HOMO orbitals. Again this shows that the ionic motion greatly alters the response of the molecule.

\begin{table*}[t]
\begin{tabular}{@{\extracolsep{0.3cm}}cccccccccc}
\hline
Polarisation  & Intensity  & Ions  & Approximation & \multicolumn{6}{c}{Harmonic Ratios} \\
              & ($\times 10^{14}$W/cm$^2$)         &      &                   & 5/7 & 7/11 & 5/3  & 11/9 & 7/9 & 11/13\\
\hline
Circular\footnote{From Reference~\cite{baer:2003}} &  3.5 & Fixed  & LDA-PW92 &3.2 & 0.9  & 11.0 & 6.8  & 6.1 &  --- \\
Circular                                           &  3.5 & Fixed  & LDA-PW92 &2.975& 5.246& ---  & ---  & --- & 0.346\\
Circular                                           &  3.5 & Moving & LDA-PW92 &3.085&4.206& ---  & ---  & --- & 0.505\\
Circular                                           &  3.5 & Fixed  & L-ADSIC    &2.768&2.603& ---  & ---  & --- & 1.184\\
Circular                                           &  3.5 & Moving & L-ADSIC    &3.149&7.660& ---  & ---  & --- & 0.537\\
\hline
Linear $\parallel$              	          &  3.5 & Moving     & L-ADSIC &1.304& 1.486& 4.546& 1.111&1.650& 5.312\\
Linear $\perp$                                    &  3.5 & Moving     & L-ADSIC &2.842& 0.193& 0.333& 0.374&0.072& 0.315\\
Linear\footnote{From Reference~\cite{hay:2000}}   & 10.0 & Experiment & ---   & --- & 1.25 & ---  & 0.50 & 0.50& --- \\
\hline
\end{tabular}
\caption{Ratios of harmonic intensities for a range of calculations using different approximations to the exchange-correlation 
functional and different descriptions of the ionic motion. For all calculations the interaction of benzene with 
a 10-cycle laser pulse having wavelength $\lambda =$ 800 nm and intensity $I = $ \intensity{3.5}{14} was studied. Blank 
entries denote situations in which one of the harmonic intensities was too small to be reliably estimated. Our results are 
compared with the previous calculations of Baer et al~\protect\cite{baer:2003}. For linear polarisation, we present the 
ratios of the given harmonics for two orientations of the molecule with the pulse. In the parallel case, the molecule lies in 
the $x-y$ plane with the pulse polarised in the $x$ direction. In 
the perpendicular case, the molecule lies in the $y-z$ plane with the pulse polarised in the $x$ direction. We compare the 
harmonic ratios with those measured experimentally by Hay et al~\protect\cite{hay:2000}.}
\label{tab:table2}
\end{table*}

\begin{figure*}[t]
\centerline{
\begin{minipage}{8.0cm}
\centerline{\includegraphics[width=8cm,viewport=21 38 565 561]{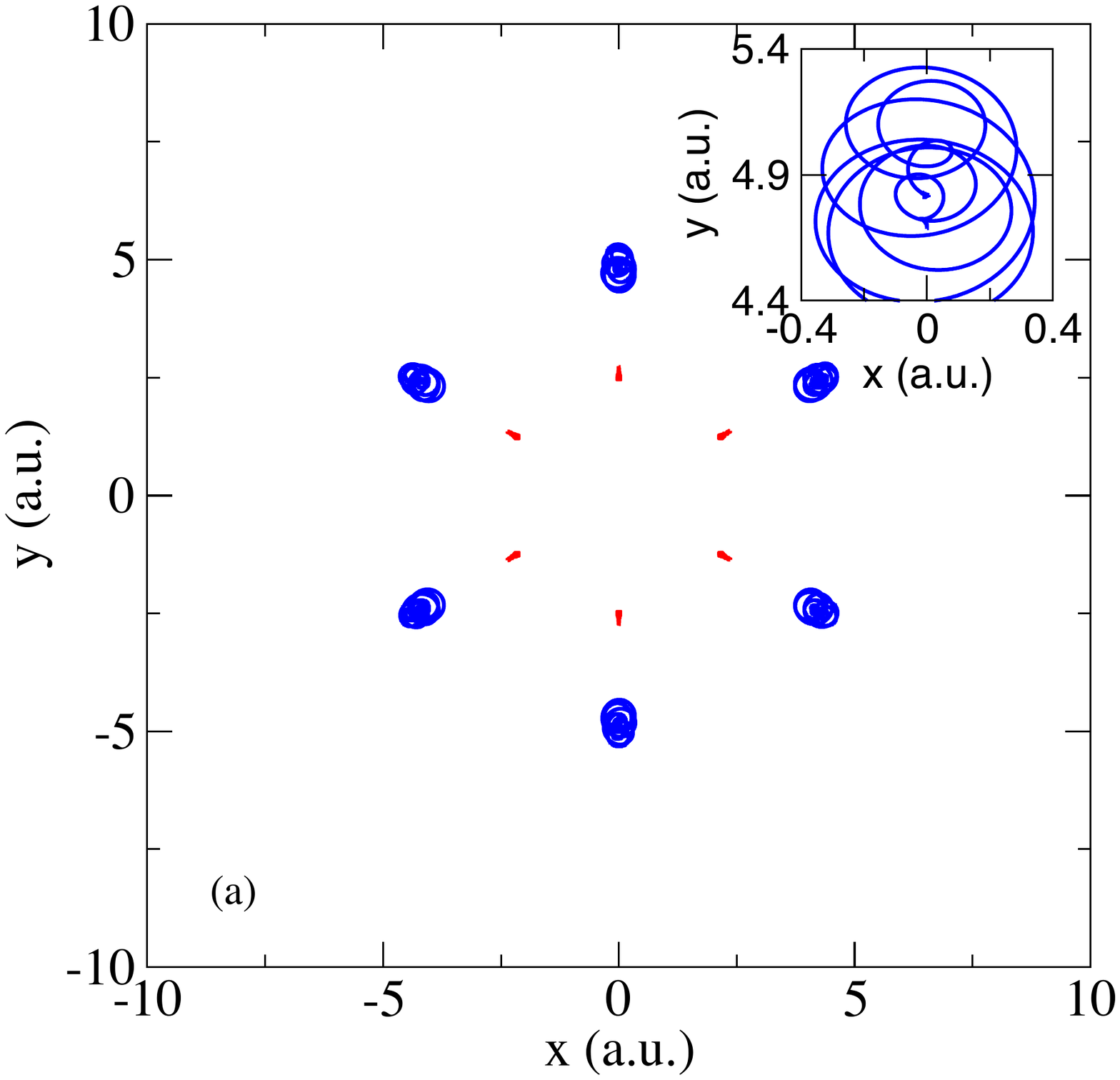}}
\end{minipage}
\hfill
\begin{minipage}{8.0cm}
\centerline{\includegraphics[width=8cm,viewport=21 38 565 561]{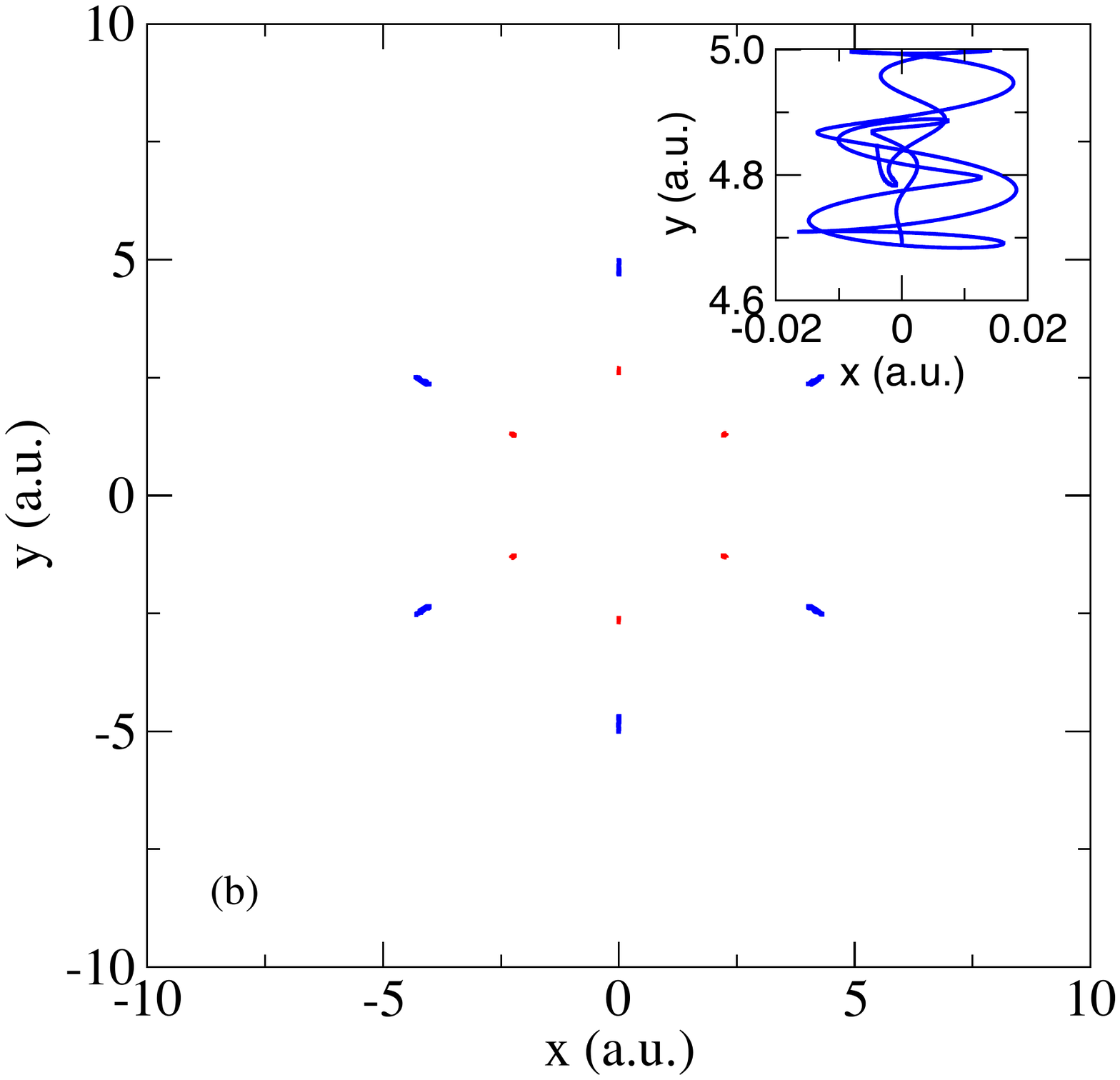}}
\end{minipage}
} 
\caption{Ion trajectories for benzene during the interaction with a 10-cycle circularly-polarised laser 
pulse of wavelength $\lambda =$ 800 nm. In (a) the peak laser intensity is $I =$ \intensity{3.5}{14} while in (b) it is
$I =$ \intensity{2.0}{14}. All calculations were performed using the L-ADSIC exchange-correlation
functional. Red trajectories correspond to carbon atoms while blue trajectories correspond to hydrogen ions. For clarity, the inset
figures show the trajectory of the top hydrogen atom in the $x-y$ plane.}
\label{fig:figure6}
\end{figure*}

In Fig.~\ref{fig:figure6} we plot the trajectories of the ions during the interaction with two different laser pulses. 
In Fig.~\ref{fig:figure6}(a) we consider the interaction of benzene with the 10-cycle laser pulse of wavelength 
$\lambda=$ 800 nm and peak intensity $I = $ \intensity{3.5}{14}. In this case we see significant response of the ions to the 
laser, even over this 26 fs timeframe. As expected, this response is greatest for the hydrogen atoms, which are observed to spiral 
around in the field. The increase in the ionic displacement during the interaction clearly increases the probability for recollisions 
to occur. In Fig.~\ref{fig:figure6}(b) we consider a calculation in which the peak laser intensity has been lowered to 
$I = $ \intensity{2.0}{14}, i.e. the same as that used in Fig.~\ref{fig:figure2}(b). In this case we see that while the hydrogen ions respond to the field,
the overall response is much less than that observed at the higher intensity. 
Additionally, the hydrogen ions now respond more randomly to the field. 
This can be understood in terms of the laser potential dominating the Coulomb potential 
at the higher intensity. At the lower intensity collective modes of motion will be present 
and so the trajectory of individual atoms may appear more random.

\subsection{Molecular response to linearly-polarised light}

\begin{figure}
\centerline{\includegraphics[width=8cm,viewport=13 42 571 493]{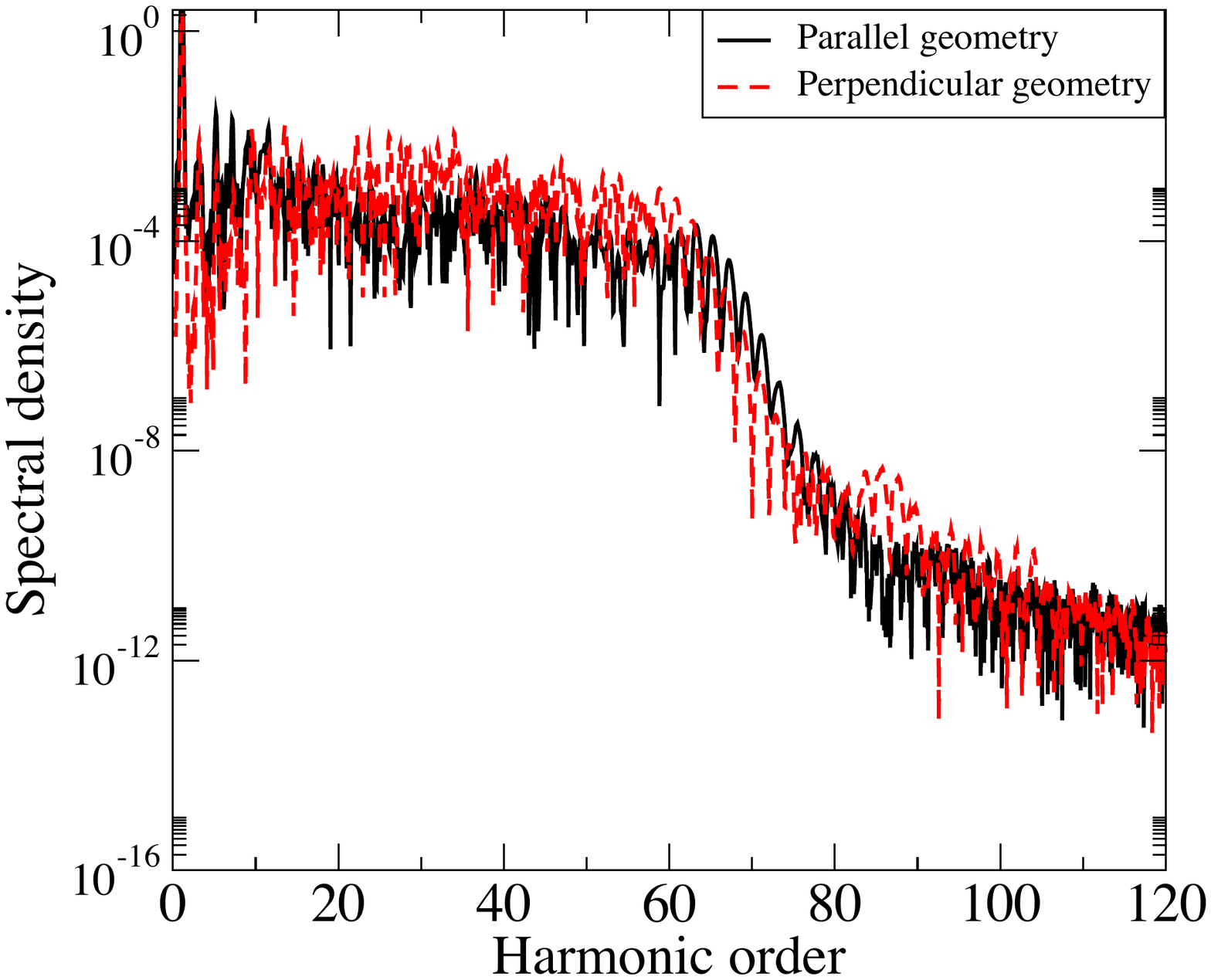}}
\caption{Calculated harmonic spectra for benzene interacting with a 10-cycle linearly polarised laser
pulse having wavelength $\lambda =$ 800 nm and peak intensity $I =$ \intensity{3.5}{14}. For both simulations the L-ADSIC 
functional was used and the ions were allowed to move. Results for two orientations of the molecule with the pulse are 
considered. In the parallel case, the molecule lies in the $x-y$ plane with the pulse polarised in the $x$ direction. 
In the perpendicular case, the molecule lies in the $y-z$ plane with the pulse polarised in the $x$ direction. }
\label{fig:figure7}
\end{figure}

At this point we have shown that the ionic motion plays a crucial role in the dynamical response of benzene to
circularly-polarised light. We would now like to investigate how the molecule responds to linearly-polarised light.
In Fig.~\ref{fig:figure2}(a) we have already shown a snapshot of the electronic density for benzene interacting with a 10-cycle 
linearly-polarised laser pulse of wavelength $\lambda =$ 800nm and peak intensity $I = $ \intensity{2.0}{14}. In that case the
laser field was aligned in the $x$ direction, i.e. in the plane of the molecule. In Fig.~\ref{fig:figure7} we show the harmonic
spectra whenever the laser intensity increases to $I = $ \intensity{3.5}{14}. Two orientations of the molecule are considered. In
the parallel orientation the molecule lies in the $x-y$ plane with the laser aligned along $x$. In the perpendicular
orientation the molecule lies in the $y-z$ plane with the laser aligned along $x$. In both simulations the L-ADSIC approximation to the
exchange-correlation functional is used and the ions are allowed to move. We plot the spectra along the laser polarisation direction, 
i.e. $S_x(\omega)$. We see that the spectra obtained are quite different to those obtained using circularly-polarised pulses. In 
particular, for linear polarisation the plateau regions are more pronounced (the cut-off harmonics in this case 
agree well with that predicted by the cut-off formula~\cite{corkum:1993,kulander:1993}). Comparing the two sets of results in 
Fig.~\ref{fig:figure7}, we see that the harmonics produced in the parallel orientation are largest for low-order harmonics 
while the plateau harmonics are greatest in the perpendicular orientation. This is in agreement with our previous LDA 
calculations that considered fixed ions~\cite{dundas:Jchemphys:2012}. 

For the lowest harmonics we can calculate their relative strengths. These are presented at the end of Table~\ref{tab:table2}. 
We see that the ratios of these harmonics vary dramatically when the molecule-laser orientation is changed. In the table, we also show
experimental ratios from Hay et al~\cite{hay:2000}. While these experimental results were obtained for a larger laser intensity, we
see that our ratios are in broad agreement. It is also clear from the results in this table that there is a large difference between 
the ratios obtained for linear and circular polarisation. This clearly illustrates the different mechanisms and selection rules that 
apply in each case.

\begin{figure*}
\centerline{
\begin{minipage}{8.0cm}
\centerline{\includegraphics[width=8cm,viewport=21 38 565 561]{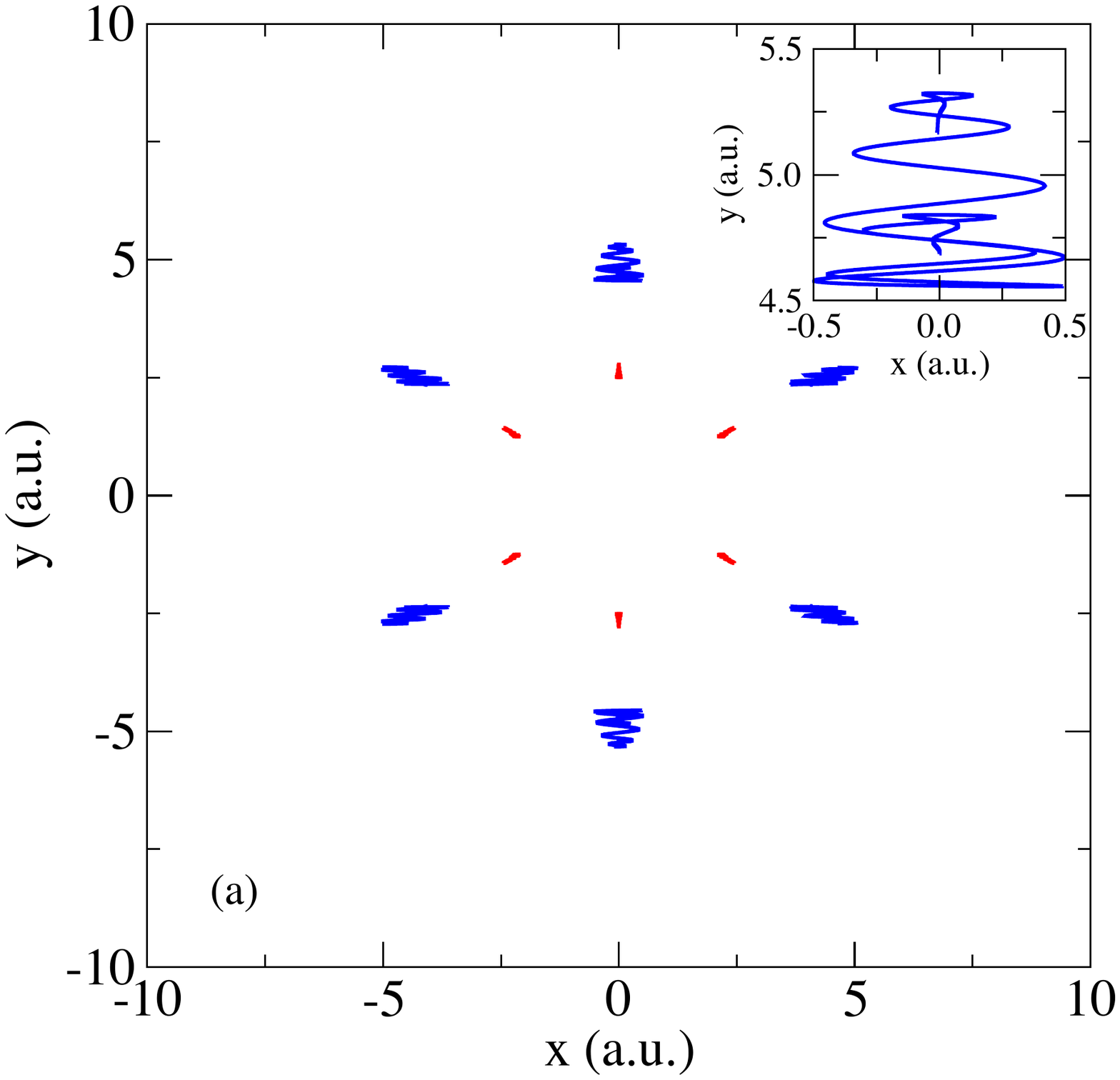}}
\end{minipage}
\hfill
\begin{minipage}{8.0cm}
\centerline{\includegraphics[width=8cm,viewport=22 37 565 561]{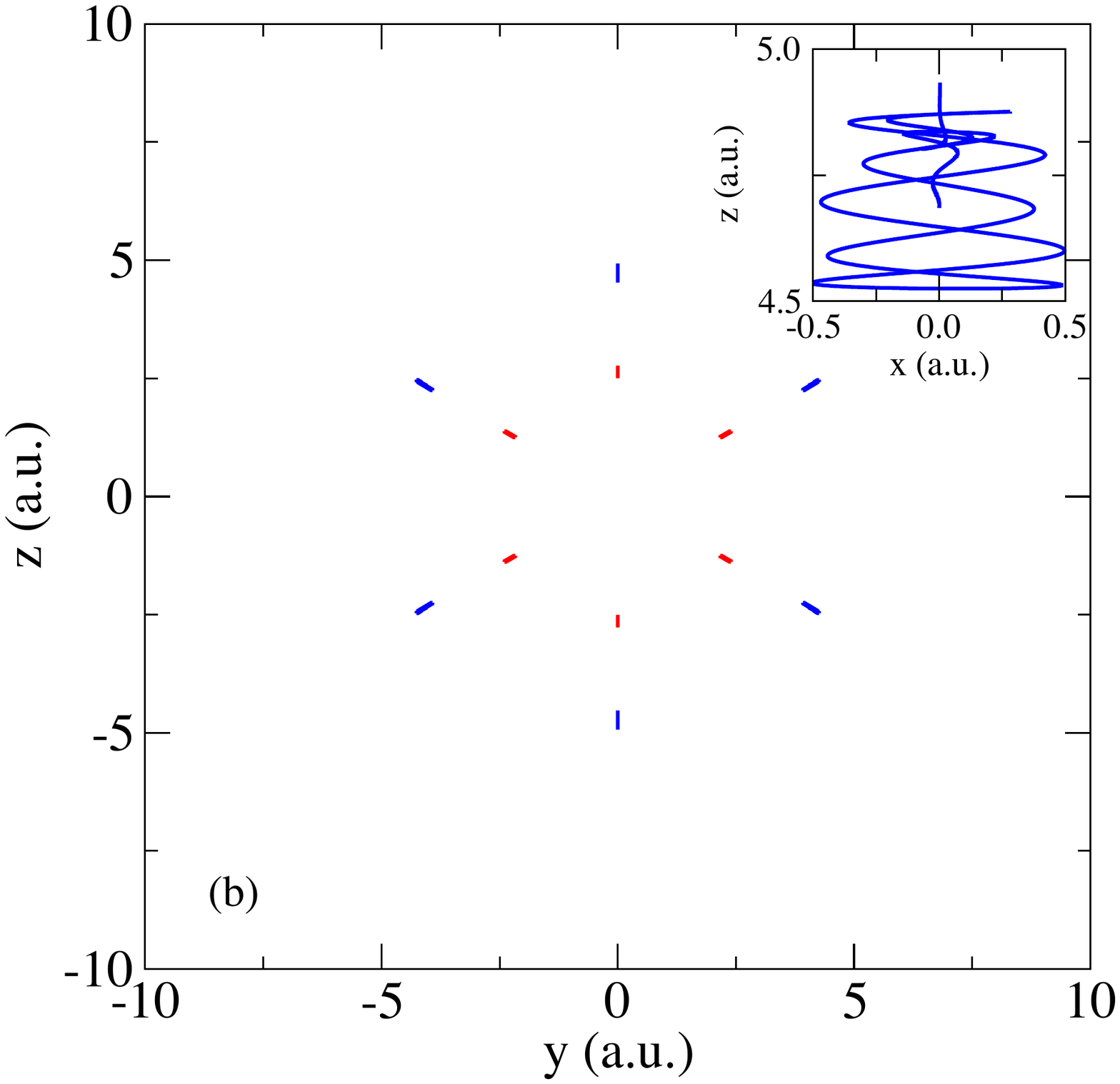}}
\end{minipage}
}

\vspace*{0.5cm}
\centerline{
\begin{minipage}{8.0cm}
\centerline{\includegraphics[width=8cm,viewport=28 37 565 498]{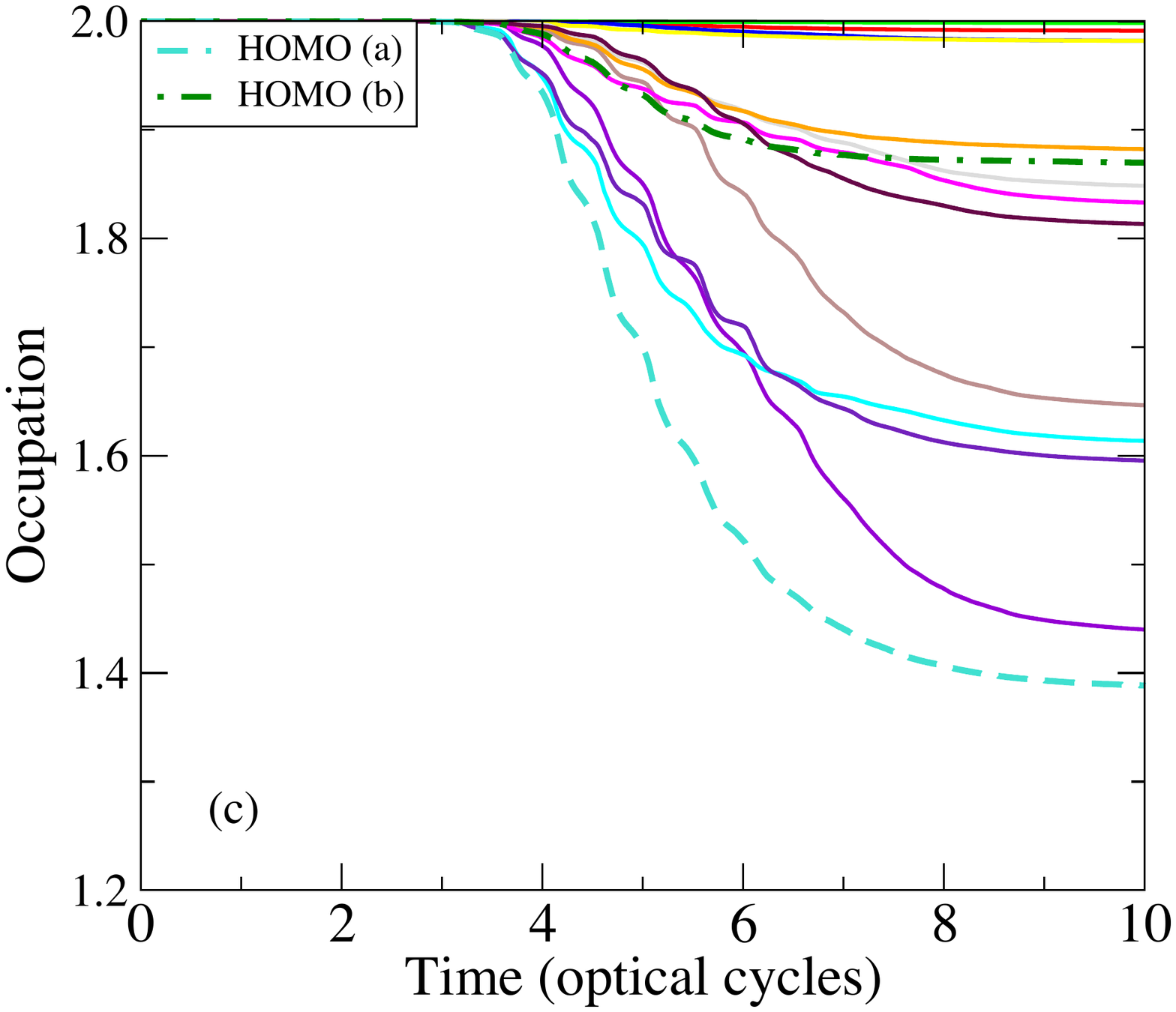}}
\end{minipage}
\hfill
\begin{minipage}{8.0cm}
\centerline{\includegraphics[width=8cm,viewport=28 37 565 498]{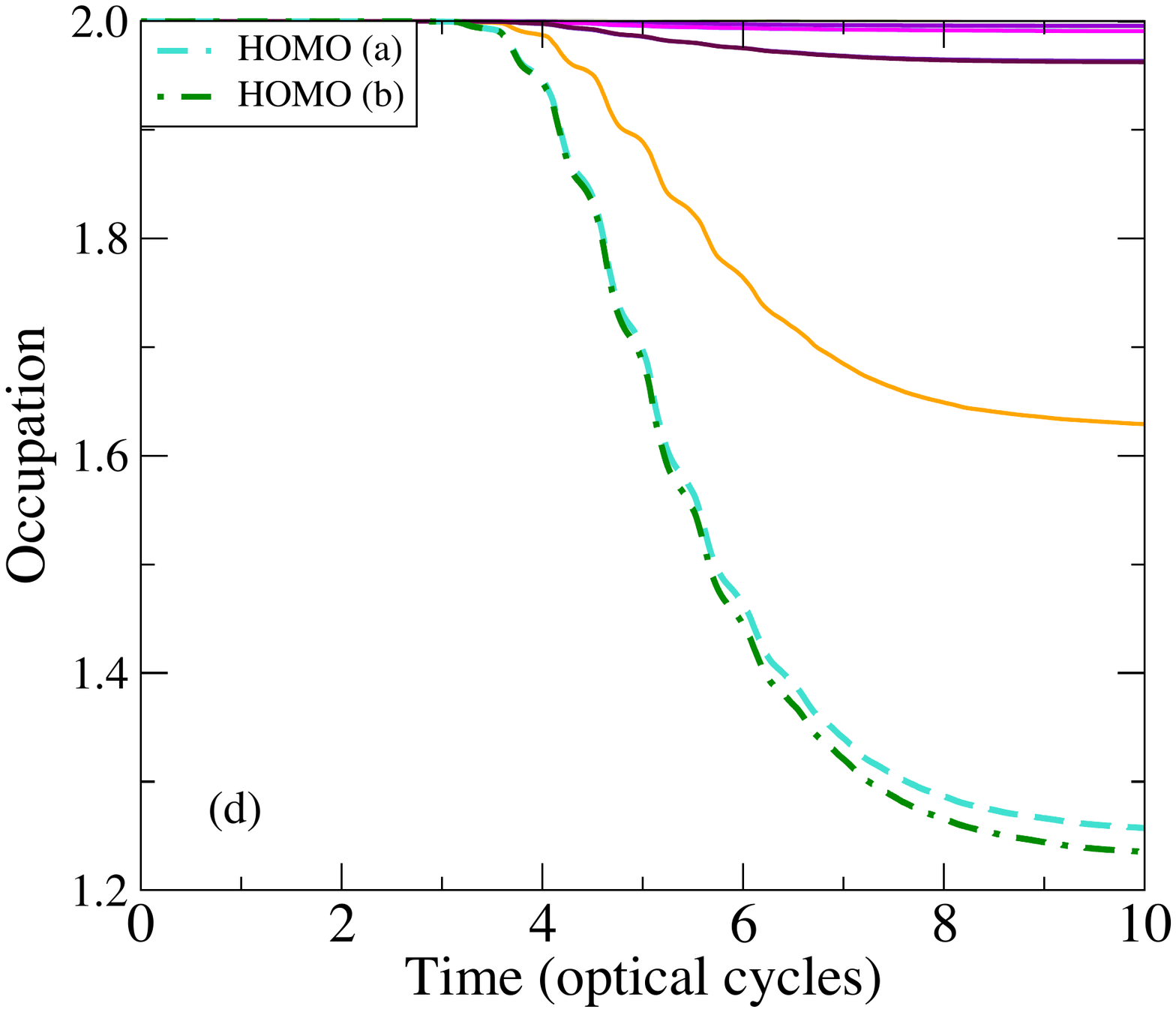}}
\end{minipage}
} 
\caption{Populations of the Kohn-Sham orbitals and ion trajectories for benzene during the interaction with a 10-cycle 
linearly-polarised laser pulse of wavelength $\lambda =$ 800 nm and peak intensity $I = $ \intensity{3.5}{14}. 
(a) and (c) respectively present the trajectories and populations for the parallel orientation between the molecule and
the field while (b) and (d) respectively present the trajectories and populations for the perpendicular orientation.
For both simulations the L-ADSIC functional was used. In the trajectory plots, red corresponds to carbon atoms 
while blue corresponds to hydrogen ions. In the population plots we only label the two forms of the 
doubly degenerate HOMO orbital [HOMO(a) and HOMO(b)] for clarity. In all cases the laser polarisation direction is aligned along the $x$
axis. The main plot of (b) presents the trajectories in the plane of the molecule (the $y-z$ plane). To show the response along the
laser polarisation axis, the inset shows the trajectory of the top hydrogen atom in the $x-z$ plane. The inset in plot
(a) shows the extent of the same hydrogen atom in the parallel orientation.}
\label{fig:figure8}
\end{figure*}

We can obtain more information about the molecular response by looking at the Kohn-Sham orbital populations and the 
trajectories of the ions.
Fig.~\ref{fig:figure8} presents the populations of the Kohn-Sham orbitals for each calculation considered in 
Fig.~\ref{fig:figure7} together with the ion trajectories. From the ion trajectories we see that the ionic
displacements are similar for both orientations considered. However, the response of the Kohn-Sham orbitals
is markedly different. In the parallel case shown in Fig.~\ref{fig:figure8}(c) we see that the 
HOMO(a) and HOMO(b) orbitals respond differently to the field and that several of the more tightly bound orbitals respond more strongly
to the pulse than the HOMO(b) orbital. In the perpendicular orientation we see that both HOMO orbitals respond almost
identically and their response is much greater than the more tightly bound orbitals. This behaviour is similar to that
observed in our earlier fixed-ion calculations using the LDA functional~\cite{dundas:Jchemphys:2012}. 

We now consider the difference in the response of benzene to linearly- and circularly-polarised laser pulses. A number of experiments
have previous considered this difference~\cite{talepour:2000,rajara:2003}. In the results of Talepour et
al~\cite{talepour:2000} it was shown that for a laser intensity of $I =$ \intensity{6.0}{14} the fragmentation patterns observed were
largely independent of the laser polarisation. Later, experiments by Rajara et al~\cite{rajara:2003} showed that for a higher laser
intensity of $I =$ \intensity{1.0}{16} the fragment yields were much greater for linear polarisation. They explained this difference
as due to electron recollisions in circularly-polarised pulses being more probable at lower laser intensities. While our results cannot be
directly compared with experiment since important propagation effects are not taken into account~\cite{jin:2011b,zhao:2011,jin:2012},
our results support
these observations. Comparing Fig.~\ref{fig:figure5}(b) with Fig.~\ref{fig:figure8}(c), we see that the population loss from the 
Kohn-Sham orbitals is greatest during interaction with the linearly-polarised pulse. Additionally, comparing 
Fig.~\ref{fig:figure6}(a) with Fig.~\ref{fig:figure8}(a) we see that the ionic response is slightly greater for both linear
polarisation. Lastly, referring to Fig.~\ref{fig:figure2} again, we see clear evidence of recollision for circular polarisation. 

\section{Conclusions}
\label{sec:conclusions}

In this paper we have studied harmonic generation in benzene using a TDDFT approach 
in which the ions were allowed to move classically while the exchange-correlation functional
included self-interaction corrections. The response to both linearly- and circularly-polarised IR laser pulses
having durations of $\sim$ 26 fs was considered. For all calculations we find that the ionic motion has a 
large effect on the harmonic response, even for such short duration pulses. In addition, we find that
the displacement of the ions is slightly greater when linearly-polarised light is used.  
Even though the probability of electron recollision in circularly-polarised pulses is lower, we still see evidence 
of recollisions in the harmonic spectra. We find that plateau harmonics (especially in the secondary plateau) are enhanced 
when the ions are allowed to move. We 
believe this is due to electrons that ionize from one ionic centre recolliding with a different ionic centre. When the ions 
are allowed to move the increased ionic response gives a greater probability for these recollisions to occur.

Baer et al~\cite{baer:2003} observed a secondary plateau in their calculations for benzene aligned in the plane of a 
circularly-polarised pulse and predicted that the length of this plateau would be similar or longer than the plateau obtained for
non-aligned benzene interacting with linearly-polarised pulses. In our calculations we have found that this is the case. However, 
we also find that the intensity of the plateau harmonics increases when the ions are allowed to move. Additionally we find that 
the population depletion from the Kohn-Sham orbitals is only slightly greater for linearly-polarised pulses than for 
circularly-polarised pulses, in agreement with experimental observations at similar intensities.
  
The method we have used is robust enough to describe general polyatomic molecules interacting with laser pulses of arbitrary 
polarisation. We can consider laser wavelengths ranging from vacuum ultraviolet (VUV) to IR. This opens up the possibility 
of carrying out simulations with attosecond pulses having both circular and linear polarisation. Therefore, we have a tool 
for studying circular dichroism in chiral molecules using HHG.

\section{Acknowledgements}
This work used the ARCHER UK National Supercomputing Service (http://www.archer.ac.uk) and has been supported by COST 
Action CM1204 (XLIC). AW acknowledges financial support through a PhD studentship funded by the UK Engineering and 
Physical Sciences Research Council.

\end{document}